\documentclass[aps,prd,letterpaper,showpacs,twocolumn,preprintnumbers,amsmath,amssymb,nofootinbib]{revtex4}
\usepackage{graphicx}

\DeclareMathAlphabet   {\mathsc}{OT1}{cmr}{m}{sc}

\def\[{\left [}
\def\]{\right ]}
\def\({\left (}
\def\){\right )}
\newcommand{\lang}{\left\langle}
\newcommand{\rang}{\right\rangle}
\newcommand{\lbr}{\left\{}
\newcommand{\rbr}{\right\}}

\newcommand{\wtd}[1]{\widetilde{#1}}

\newcommand{\GeV}      {~\mathrm{GeV}}
\newcommand{\TeV}      {~\mathrm{TeV}}

\newcommand{\UV}       {\mathsc{uv}}

\newcommand{\GUT}      {\mathsc{gut}}

\newcommand{\SUSY}     {\mathsc{susy}}

\newcommand{\Ohsq}{\Omega_{\chi} {\rm h}^2}

\newcommand{\order}{\mathcal{O}}

\newcommand{\gappeq}{\mathrel{\rlap {\raise.5ex\hbox{$>$}}
{\lower.5ex\hbox{$\sim$}}}}
\newcommand{\lappeq}{\mathrel{\rlap{\raise.5ex\hbox{$<$}}
{\lower.5ex\hbox{$\sim$}}}}

\begin{document}

\preprint{LBNL-51698} \preprint{UCB-PTH-02/50}
\preprint{MCTP-02-53}

\title{Relic Neutralino Densities and Detection Rates \\ with Nonuniversal
Gaugino Masses}

\author{Andreas Birkedal-Hansen}
\affiliation{Department of Physics, University of California, Berkeley, CA 94720, USA and \\
Theoretical Physics Group, Lawrence Berkeley National Laboratory, Berkeley, CA 94720, USA}

\author{Brent D. Nelson}
\affiliation{Michigan Center for Theoretical Physics, University of Michigan, Ann Arbor, MI 48109, USA}

\date{\today}

\begin{abstract}
We extend previous analyses on the interplay between
nonuniversalities in the gaugino mass sector and the thermal relic
densities of LSP neutralinos, in particular to the case of
moderate to large $\tan\beta$. We introduce a set of parameters
that generalizes the standard unified scenario to cover the
complete allowed parameter space in the gaugino mass sector. We
discuss the physical significance of the cosmologically preferred
degree of degeneracy between charginos and the LSP and study the
effect this degree of degeneracy has on the prospects for direct
detection of relic neutralinos in the next round of dark matter
detection experiments. Lastly, we compare the fine tuning required
to achieve a satisfactory relic density with the case of universal
gaugino masses, as in minimal supergravity, and find it to be of a
similar magnitude. The sensitivity of quantifiable measures of
fine-tuning on such factors as the gluino mass and top and bottom
masses is also examined.
\end{abstract}

\pacs{12.60.Jv,04.65.+e,14.80.Ly,95.35.+d}

\maketitle

\section*{Introduction}

The perplexing question of the nature of the missing non-baryonic
matter in the universe continues to be an active area of research.
In supersymmetry, the lightest supersymmetric particle (LSP) tends
to be a neutral gaugino, which is stable if the assumption of
R-parity conservation is imposed, and its weak-scale
self-interaction rate in the early universe make it an attractive
candidate to fill this role. Not surprisingly, then, relic
neutralinos continue to be the favored candidate for the missing
matter density of the universe.

Most phenomenological studies of the parameter space in
supersymmetric models that give rise to the cosmologically
preferred thermal relic density~\cite{BaOsPeSt99,Fr00} of $0.1
\leq \Omega_{\chi} {\rm h}^2 \leq 0.3$ focus on the minimal
supergravity model (mSUGRA) with its unified gaugino mass
$m_{1/2}$ and unified scalar mass $m_0$. In such constrained
models of the MSSM\footnote{The class of models which we refer to
as mSUGRA is also commonly known as the constrained MSSM, or
``CMSSM,'' in the literature.} a typical point in the $\( m_{1/2},
\; m_0\)$ plane that would otherwise be quite ``natural'' from the
point-of-view of solving the hierarchy problem of the electroweak
sector tends to produce far {\em too much} relic neutralino
density in the early universe to be compatible with the known age
of the universe. A sizable portion of the parameter space that
does not come into conflict with the bound $\Omega_{\chi} {\rm
h}^2 \alt 0.3$ is instead in conflict with the bound on the
lightest CP-even Higgs mass coming from LEP, particularly for low
values of the parameter $\tan\beta$.

This disturbing situation was pointed out in our previous
study~\cite{BiNe01}, where it was observed that significant
improvement can be achieved if the assumption of universal gaugino
masses is relaxed. The problematic behavior of the mSUGRA model is
a result of the overwhelming B-ino content of the LSP and the
large mass difference between the LSP and the next lightest
supersymmetric particle (NLSP) which holds over most of the
parameter space in such models. Combined with the need for a large
universal gaugino mass to accommodate a Higgs mass of $115 \GeV$
these properties lead to an excess of relic B-inos outside of
certain special areas where resonant annihilation or
coannihilation processes dominate. When this assumption of gaugino
mass universality is not valid -- in particular when the W-ino
mass $M_2$ is less than the B-ino mass $M_1$ at the initial
supersymmetry breaking scale $\Lambda_{\UV}$ -- the increased
W-ino content of the LSP and reduced mass difference $\Delta m
\equiv m_{\chi^{\pm}} - m_{\chi^0_1}$ between the LSP neutralino
$\chi^0_1$ and NLSP chargino $\chi^{\pm}$ can have dramatic
effects on the depletion of relic neutralinos in the early
universe. This will open up new areas of MSSM parameter space with
acceptable relic densities. Quite apart from this improvement in
the dark matter arena, such nonuniversalities in the gaugino
sector are a common feature of string-derived effective theories.
It was further suggested recently~\cite{KaLyNeWa02} that
nonuniversalities in the gaugino sector may also play an important
role in ameliorating the apparent fine-tuning of the MSSM required
in the electroweak sector as well.

In the current work we utilize newly-available calculation tools
to extend the analysis to include the important large-$\tan\beta$
region which has been of great interest recently in the context of
minimal supergravity models. We pay particular attention to the
behavior of that area of parameter space in which resonant
annihilation through heavy Higgs states occurs when gaugino mass
nonuniversality is introduced.  We find that the viable parameter
space is enlarged, with significant regions allowing for heavy
scalars and relatively light gauginos. The physical significance
behind this new parameter region is investigated and the
sensitivity of these results to the value of the gluino mass is
explored. This analysis forms the bulk of
Section~\ref{sec:density}.

Just as changing the relative size of the gaugino mass parameters
can have large effects on the number and nature of coannihilation
channels available to the relic LSP in the early universe, so too
can it have profound impacts on the interaction rates of surviving
relics with current dark matter detectors. In
Section~\ref{sec:detection} we investigate the prospects for
direction detection of relic neutralinos in our local halo in the
next round of future detectors. While nonuniversalities that give
rise to a relic density in the preferred range typically generate
interaction cross-sections too small for current detectors such as
DAMA or CDMS, we find a significant region of the preferred
parameter space will be probed by the next generation of Germanium
and Xenon-based detectors.

Lastly, in an effort to quantify the difference between the mSUGRA
paradigm and the case with nonuniversal gaugino masses, we adopt a
fine-tuning sensitivity parameter similar to that of Ellis and
Olive~\cite{ElOl01}.  We calculate the sensitivity of the
neutralino relic density to small variations in the high scale
parameters, including input values such as the top and bottom
quark masses. In Section~\ref{sec:tuning} we compare the typical
amounts of tuning in universal and nonuniversal scenarios, and
comment upon the implications, before concluding.

\section{\label{sec:density}Relic Density of LSP Neutralinos}

We are interested in studying the present density of neutralinos
assuming they were once in thermal equilibrium with radiation in
the early universe. The lightest such neutralino will be the LSP
in all models we will consider. The mass and couplings of the
neutralino sector at tree level are determined by the mass matrix
\begin{widetext}
\begin{equation}
\(\begin{array}{cccc}
M_{1} & 0 & -\sin\theta_{W}\cos\beta M_{Z} &
\sin\theta_{W}\sin\beta M_{Z} \\
0 & M_{2} & \cos\theta_{W}\cos\beta M_{Z}& -
\cos\theta_{W}\sin\beta M_{Z}\\
-\sin\theta_{W}\cos\beta M_{Z} & \cos\theta_{W}\cos\beta M_{Z} & 0
& -\mu \\
\sin\theta_{W}\sin\beta M_{Z} & -\cos\theta_{W} \sin\beta M_{Z} &
-\mu & 0 \end{array}\),
\label{neutmatrix} \end{equation}
\end{widetext}
%
%
%
%
%
which is written in the $(\wtd{B}, \wtd{W}, \wtd{H}^{0}_{d},
\wtd{H}^{0}_{u})$ basis, where $\wtd{B}$ represents the B-ino,
$\wtd{W}$ represents the neutral W-ino and $\wtd{H}^{0}_{d}$ and
$\wtd{H}^{0}_{u}$ are the down-type and up-type Higgsinos,
respectively. The parameters $M_1$ and $M_2$ in~(\ref{neutmatrix})
are the soft supersymmetry-breaking gaugino masses for the B-ino
and W-ino, respectively while $\mu$ is the supersymmetric Higgsino
mass parameter. The ratio of Higgs vevs is given by the parameter
$\tan\beta=v_u/v_d$ with $v_u$ the vacuum expectation value of the
up-type Higgs and $v_d$ the vacuum expectation value of the
down-type Higgs. We will assume that all of the parameters
in~(\ref{neutmatrix}) are real for this analysis.\footnote{See,
however, Ref.~\onlinecite{BrChKa01} for the implications on
neutralino relic densities when this assumption is relaxed.}

The content of the lightest neutralino can be parameterized by
writing the LSP as
\begin{equation}
\chi^{0}_{1} = N_{11} \wtd{B} + N_{12} \wtd{W} + N_{13}
\wtd{H}^{0}_{d} + N_{14} \wtd{H}^{0}_{u},
\label{LSPcontent} \end{equation}
with the normalization $N_{11}^2+N_{12}^2+N_{13}^2+N_{14}^2=1$.
When the parameter $\mu$ is much larger than the Z-boson mass it
is clear that the proportion of W-ino content to B-ino content in
the LSP is determined by the ratio of their soft masses, which we
denote by $r\equiv M_2/M_1$. In mSUGRA this parameter is defined
to be $r=1$ at some high-energy input scale, usually taken to be
the GUT scale $\Lambda_{\UV} = \Lambda_{\GUT} = 2 \times 10^{16}
\GeV$. Upon renormalization group (RG) evolution of these
parameters to the electroweak scale this implies $M_1 \simeq
\frac{1}{2} M_2$ and the lightest eigenvalue of~(\ref{neutmatrix})
is thus overwhelmingly B-ino in content ($N_{11} \simeq 1$). In a
more general scenario the neutralino sector would be defined by
four parameters at the electroweak scale: $\lbr \tan\beta,\; \mu,
\; r, \; M_{1} \rbr$.

In practice the magnitude of the $\mu$ parameter is determined at
the electroweak scale by enforcing proper electroweak symmetry
breaking (EWSB). Here this will be performed by computing the
complete one loop corrected effective potential $V_{\rm
1-loop}=V_{\rm tree} + \Delta V_{\rm rad}$. The correction $\Delta
V_{\rm rad}$ includes radiative effects from the entire
superpartner spectrum and thus involves all of the parameters of
the soft supersymmetry-breaking Lagrangian. The effective
$\mu$-term $\bar{\mu}$ is calculated from the EWSB condition
\begin{equation}
{\bar{\mu}}^{2}=\frac{\(m_{H_d}^{2}+\delta m_{H_d}^{2}\) -
  \(m_{H_u}^{2}+\delta m_{H_u}^{2}\) \tan{\beta}}{\tan^{2}{\beta}-1}
-\frac{1}{2} M_{Z}^{2} ,
\label{radmuterm} \end{equation}
where $\delta m_{H_u}$ and $\delta m_{H_d}$ are the second
derivatives of the radiative corrections $\Delta V_{\rm rad}$ with
respect to the up-type and down-type Higgs scalar fields,
respectively. The sign of the $\mu$ parameter is not determined
and left as a free parameter.

Determination of $\mu$ through~(\ref{radmuterm}), as well as the
determination of the one loop radiative corrections to the
neutralino mass matrix~(\ref{neutmatrix}), involve the remainder
of the supersymmetry breaking soft terms. These soft terms will
also be necessary to translate low energy values
in~(\ref{neutmatrix}) into high energy input values by solving the
renormalization group equations (RGEs). We wish to isolate the
effects of gaugino mass nonuniversality so we choose to consider a
universal scalar mass $m_0$ and universal trilinear A-term $A_0$
as in minimal supergravity. We are thus led to consider the
seven-dimensional parameter space we refer to as {\em rSUGRA}
given by
\begin{equation}
\lbr \tan\beta,\; {\rm sgn}(\mu), \; r, \; M_{1}, \; M_{3}, \;
m_0, \; A_0 \rbr .
\label{rSUGRA} \end{equation}
The effect of the trilinear coupling $A_0$ is negligible for the
physics we are considering here so we will set $A_0 = 0$
throughout. We will also restrict our attention to positive $\mu$.
This is very similar to the parameter space originally studied by
Mizuta, Ng and Yamaguchi~\cite{MiNgYa93}, where a trio of specific
values for the parameter $r$ were investigated. The effect of
gaugino mass nonuniversality on neutralino dark matter in certain
specific models has also been computed recently in
Ref.~\onlinecite{CoNa01}~and~\onlinecite{BeNeOr02}. The latter
appeared as we completed this analysis and we will comment on
areas of agreement with that work during the following sections.

\begin{figure}
  \includegraphics[scale=0.45]{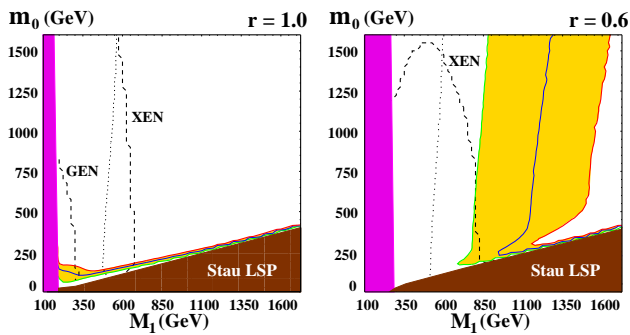}
  \caption{\label{fig:tan5} {\bf Neutralino Relic Densities for
            $\tan\beta = 5$}. The preferred neutralino relic density regions
            (light shading) are displayed for $r=1$ (left panel)
            and for $r=0.6$ (right panel).
            Dark shaded regions in the lower right are excluded due to a charged LSP.
            Medium shaded regions on the left are excluded by LEP bounds on the
            chargino mass.
            The Higgs mass contour of $m_h = 113\GeV$ is given by the dotted vertical
            line.
            The dashed contours are a rough estimate of the direct
            detection reach of the GENIUS (``GEN'') and XENON (``XEN'') experiments.}
\end{figure}

\begin{figure}
  \includegraphics[scale=0.45]{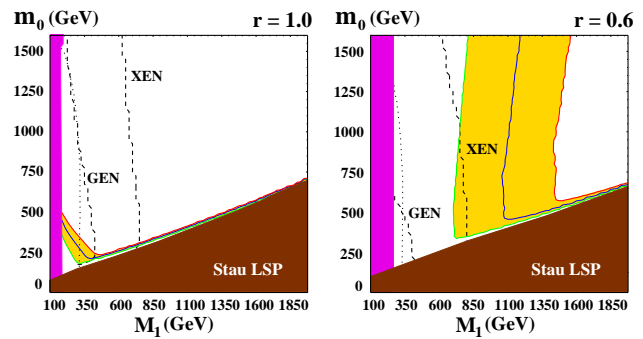}
  \caption{\label{fig:tan35} {\bf Neutralino Relic Densities for
            $\tan\beta = 35$}.  The preferred neutralino relic density regions
            (light shading) are displayed for $r=1$ (left panel)
            and for $r=0.6$ (right panel).
            Dark shaded regions in the lower right are excluded due to a charged LSP.
            Medium shaded regions on the left are excluded by LEP bounds on the
            chargino mass.
            The Higgs mass contour of $m_h = 113\GeV$ is given by the dotted vertical
            line.
            The dashed contours are a rough estimate of the direct
            detection reach of the GENIUS (``GEN'') and XENON (``XEN'') experiments.}
\end{figure}

Translation of the parameter set~(\ref{rSUGRA}) from their values
at the high scale (we will assume $\Lambda_{\UV} =
\Lambda_{\GUT})$ is performed using the Fortran code {\tt
SuSpect}~\cite{SUSPECT} which solves the RGEs at two loops between
the scale $\Lambda_{\GUT}$ and $M_Z$ through an iterative
procedure that takes into account proper thresholding of
superpartners. Proper EWSB is imposed and the one loop corrected
$\mu$ parameter is determined. The complete superpartner spectrum
is calculated including one loop radiative corrections.

The soft supersymmetry breaking parameters at the weak scale are
then passed to the C code {\tt micrOMEGAs}~\cite{MICRO} to perform
the relic density calculation. This consists of computing a
thermally averaged annihilation cross section times LSP velocity
$\lang \sigma v \rang_{\rm ann}$ and freeze-out temperature $x_{F}
= T_{F}/m_{\chi^{0}_{1}}$ for the LSP
$\chi_{1}^{0}$~\cite{OlSr91,GoGe91,DrNo93,JuKaGr96,BaKa98}.
Freeze-out occurs when the relic density of the LSP no longer
tracks the density of relativistic degrees of freedom in the
universe but is instead nearly constant. The freeze-out
temperature is defined in {\tt micrOMEGAs} as the temperature at
which the relic density of $\chi_{1}^{0}$ becomes roughly twice
the value that would obtain in equilibrium. The power of {\tt
micrOMEGAs} lies in its inclusion not just of direct annihilation
processes involving two neutralinos such as
$\chi_{1}^{0}\chi_{1}^{0} \to \ell^{+} \ell^{-}$, but also of all
possible coannihilation channels involving the relic LSP such as
$\chi_{1}^{0} \chi^{\pm} \to W^{\pm} Z$ and $\chi_{1}^{0}
\tilde{\tau} \to \gamma \tau$. These processes are known to be
crucial to a proper calculation of the relic density whenever the
masses of the LSP and of the coannihilating particle are nearly
degenerate~\cite{GrSe91,EdGo97,ElFaOlSr00,ElOlSa01,BaBaBe02,NiRoRu02a,BiJe02,ElFaOlSa02}.

\subsection{\label{sec:M3M1}The case of gluino/b-ino equality}

As mentioned in the introduction, the typical mSUGRA analysis has
the luxury of displaying results in the $\lbr m_{1/2}, \; m_0\rbr$
plane for a given value of $\tan\beta$. Our parameter set given
in~(\ref{rSUGRA}) introduces two new degrees of freedom for the
gaugino sector, so we will begin our analysis by exhibiting the
neutralino relic density in the $\( M_1, \; m_0\)$ plane for
specific values of $r$ and $\tan\beta$. In order to make contact
with the familiar results of recent mSUGRA
studies~\cite{ElFaGaOl00,FeMaWi00,BaKa01,DjDrKn01,BaBaBeMiTaWa02}
we will set $M_3 = M_1$ at the initial GUT scale in this
subsection. Since the bulk of the RGE effects of a given value of
$m_{1/2}$ in unified models is related to the value of $M_3$, this
choice most closely mimics the RGE behavior of the universal
mSUGRA scenario.

\begin{figure}
  \includegraphics[scale=0.45]{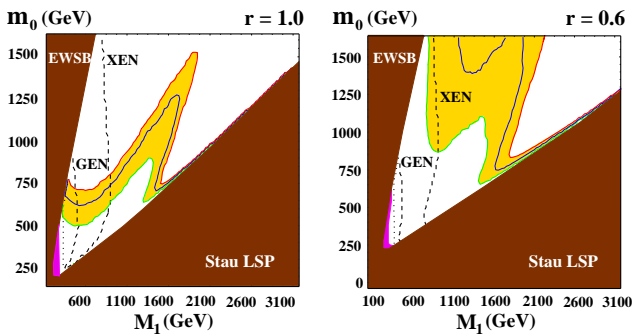}
  \caption{\label{fig:tan50} {\bf Neutralino Relic Densities for
            $\tan\beta = 50$}.  The preferred neutralino relic density regions
            (light shading) are displayed for $r=1$ (left panel)
            and for $r=0.6$ (right panel).
            Dark shaded regions in the lower right are excluded due to a charged LSP,
            while those in the upper left are excluded due to improper EWSB.
            Medium shaded regions in the lower left are excluded by LEP bounds on the
            chargino mass.
            The Higgs mass contour of $m_h = 113\GeV$ is given by the dotted vertical
            line.
            The dashed contours are a rough estimate of the direct
            detection reach of the GENIUS (``GEN'') and XENON (``XEN'') experiments.}
\end{figure}

The results of this analysis are shown in
Figures~\ref{fig:tan5},~\ref{fig:tan35} and~\ref{fig:tan50} for
values of $\tan\beta$ of 5, 35 and 50 respectively. In each of
these figures, we display available parameter space in the
$\left(M_1, m_{0}\right)$ plane, with $M_1$ and $m_0$ being the
soft supersymmetry breaking parameters at the initial (GUT) scale,
for $r=1$ and $r=0.6$. The former value recovers the case of
minimal supergravity, while the latter value was observed in
Ref.~\onlinecite{BiNe01} to be the area where nonuniversality in
the gaugino mass sector reaches a critical value for which the
preferred relic density becomes roughly independent of scalar mass
values. The lightly shaded area in each region is the area of the
parameter space preferred by cosmology, with $0.1 \leq
\Omega_{\chi} {\rm h}^2 \leq 0.3$. The left-most contour (green)
contour bordering this region is the contour of $\Omega_{\chi}
{\rm h}^2=0.1$ while the right-most (red) contour bordering the
region is the contour of $\Ohsq=0.3$. Recent analysis of all
applicable cosmological data~\cite{MeSi02} suggests a narrower
region for lower values of $\Ohsq$ may be preferred by
observational data. We therefore provide the (blue) contour for
$\Omega_{\chi} {\rm h}^2=0.2$, within the preferred region, as a
guide to this more restrictive parameter space.

The darkly shaded areas in each plot are ruled out by requiring
the lightest neutralino to be the lightest superpartner (the
region in the lower right) or by requiring that correct
electroweak symmetry breaking occurs (the region in the upper left
on some plots). While it has become fashionable for dark matter
calculational tools to include computations of indirect
constraints on supersymmetry such as ${\rm BR}(b\to s \gamma)$ and
$\delta (a_{\mu})|_{\SUSY}$, we have chosen not to include such
information in our figures to reduce clutter.\footnote{It should
be noted that {\tt micrOMEGAs} also computes such quantities, as
well as applying direct collider limits.} We have checked,
however, that with our choice of $\mu >0$ the conservative
1-$\sigma$ lower limit on the branching ratio for inclusive decays
${\rm BR}(b\to s \gamma) \geq 2.2 \times 10^{-4}$ coming from
CLEO~\cite{CLEO} is satisfied for all of the parameter choices we
exhibit. So too is the conservative upper bound~\cite{MaWe02} on
the SUSY contribution to the muon anomalous magnetic moment
$\delta (a_{\mu})|_{\SUSY} \leq 90 \times 10^{-10}$~\cite{MUON}
for all but the slimmest region in the lower left corner of these
figures -- a region irrelevant to the parameter region we are
exploring in this paper.
Two key experimental constraints which we do display are the Higgs
and chargino mass limits from LEP. Regions excluded by the LEP
limit on the mass of the lightest chargino $m_{\chi_{1}^{\pm}}
\geq 103 \GeV$~\cite{LEPchi1,LEPchi2} are indicated by the medium
(magenta) shaded regions on the left of the plots. We have chosen
to be conservative and indicate the contour of $m_h = 113 \GeV$
for the lightest CP-even Higgs mass by the dotted line as an
indication of the area excluded by LEP~\cite{LEPhiggs}.

Finally, the estimated detection reach of the proposed Germanium
detector GENIUS, and the next-generation liquid Xenon-based XENON
detector, are indicated by the heavy dashed lines labeled ``GEN''
and ``XEN,'' respectively. These are obtained by taking the
estimated minimal cross section for neutralino-nucleon elastic
scattering, $\sigma_{\chi p}$, capable of producing a signal in
GENIUS~\cite{GENIUS} and XENON~\cite{XENON}, as a function of the
LSP mass $m_{\chi_1^0}$. This is translated into a contour in the
$\(M_1,\; m_0\)$ plane and normalized for a neutralino relic
density of $\Ohsq = 0.1$. We will discuss these contours in more
detail in Section~\ref{sec:detection}.

We begin with the cases for low to moderate $\tan\beta$ in
Figures~\ref{fig:tan5} and~\ref{fig:tan35}, which display similar
features. For $r=1$ there is very little parameter space available
for which the Higgs exclusion bound is not violated, particularly
for low $\tan\beta$. One is required to be in the thin
coannihilation tail where the lightest stau is almost exactly
degenerate with the lightest neutralino. This situation is
well-known in mSUGRA (the $r\to 1$ limit of rSUGRA). However, once
$r$ is lowered towards the critical value of $r=0.6$ we see that a
wide plume opens up, independent of $m_{0}$, that is perfectly
consistent with all experimental constraints arising from
accelerators. For both low and moderate $\tan\beta$ this preferred
region for $r=0.6$ arises at relatively large values of $M_1$,
implying relatively large values of the gluino mass, and thus
occurs in areas of the parameter space where the Higgs mass can
easily exceed the LEP limit.

For extremely large values of $\tan\beta$, as
Figure~\ref{fig:tan50} demonstrates, the Higgs mass constraint is
satisfied throughout most of the parameter space. Here resonant
annihilation of neutralinos can proceed through an s-channel
exchange of neutral heavy Higgs states, particularly the
pseudoscalar. This generates an increase in the cosmologically
preferred region, even for $r=1$. The large size of this region is
a direct result of the large width of the CP-odd Higgs mass in
this high $\tan\beta$ regime and has generated much recent
interest~\cite{BaBaBe02,BaBaBeMiTaWa02,ElNaOl01,ElFaGaOlSr01,NiRoRu02,LaSp02}.
The position, and indeed even the existence, of this large region
for any given value of $\tan\beta$ is very sensitive to the Yukawa
couplings of the bottom and top quarks~\cite{NiRoRu02}. The pole
masses we used in {\tt SuSpect} for this analysis were $m_{t}^{\rm
pole} = 175.0 \GeV$ and $m_{b}^{\rm pole} = 4.9 \GeV$. This choice
is slightly higher than the default setting of $m_{b}^{\rm pole} =
4.6 \GeV$ in {\tt SuSpect}, but we find this value allows us to
match our results most closely with those of Refs.
~\onlinecite{ElNaOl01,ElFaGaOlSr01,NiRoRu02} for $\tan\beta=50$.
This should be seen as evidence of the sensitivity of high
$\tan\beta$ results to the subtleties of one's RGE code and how
one treats the NLO corrections to the Higgs sector.\footnote{For
an analysis of mSUGRA parameter space involving the default {\tt
SuSpect} settings, one may consult Ref.~\onlinecite{DjDrKn01}.}
From the right panel of Figure~\ref{fig:tan50} it is clear that
the presence of this Higgs pole region continues to distort the
allowed parameter space as we increase the degree of
nonuniversality in the B-ino/W-ino system, but the sensitivity of
this allowed space to Yukawa couplings is greatly reduced. We will
return to this issue in Section~\ref{sec:tuning} below.

\begin{figure}
  \includegraphics[scale=0.45]{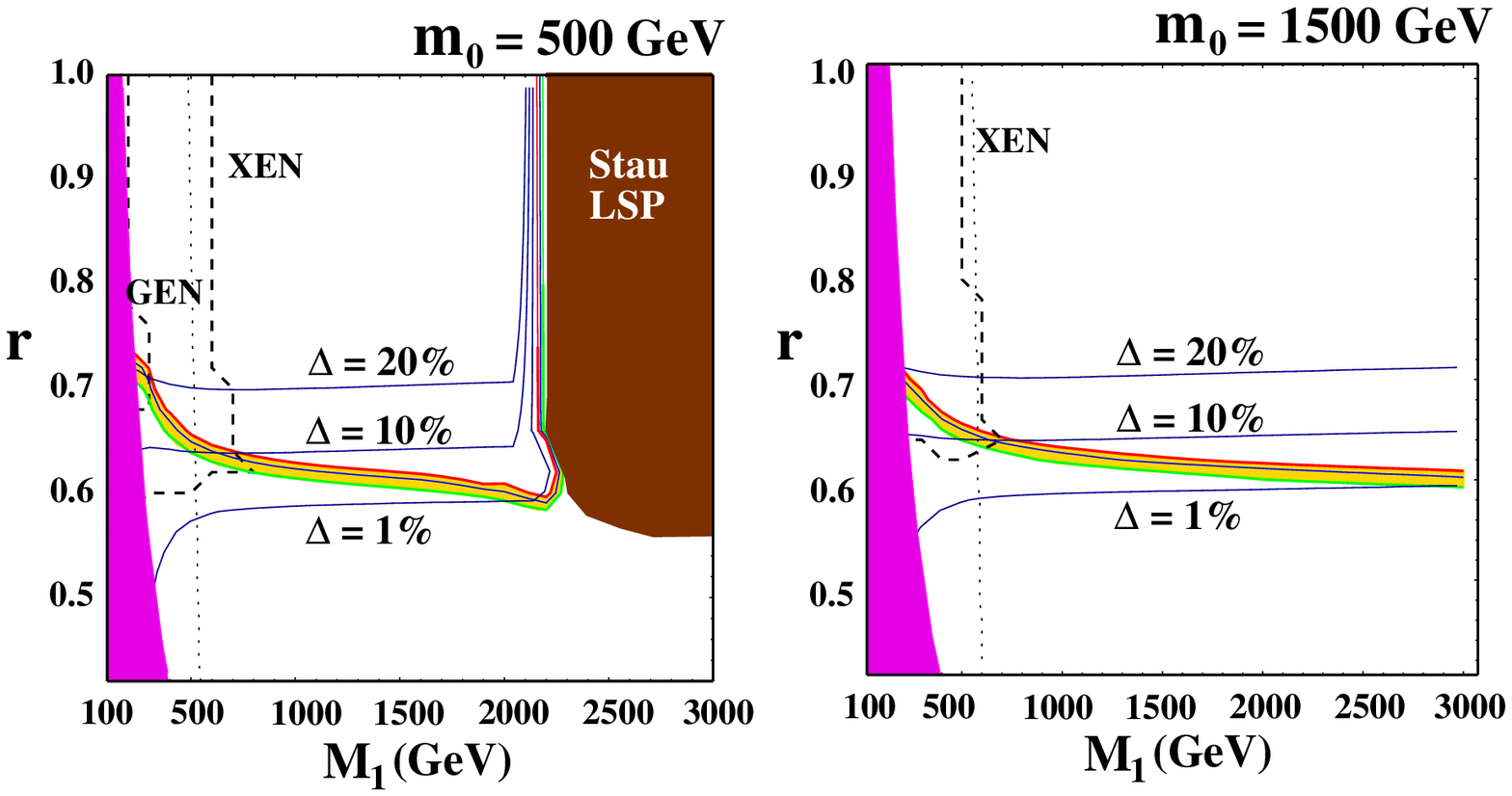}
  \includegraphics[scale=0.45]{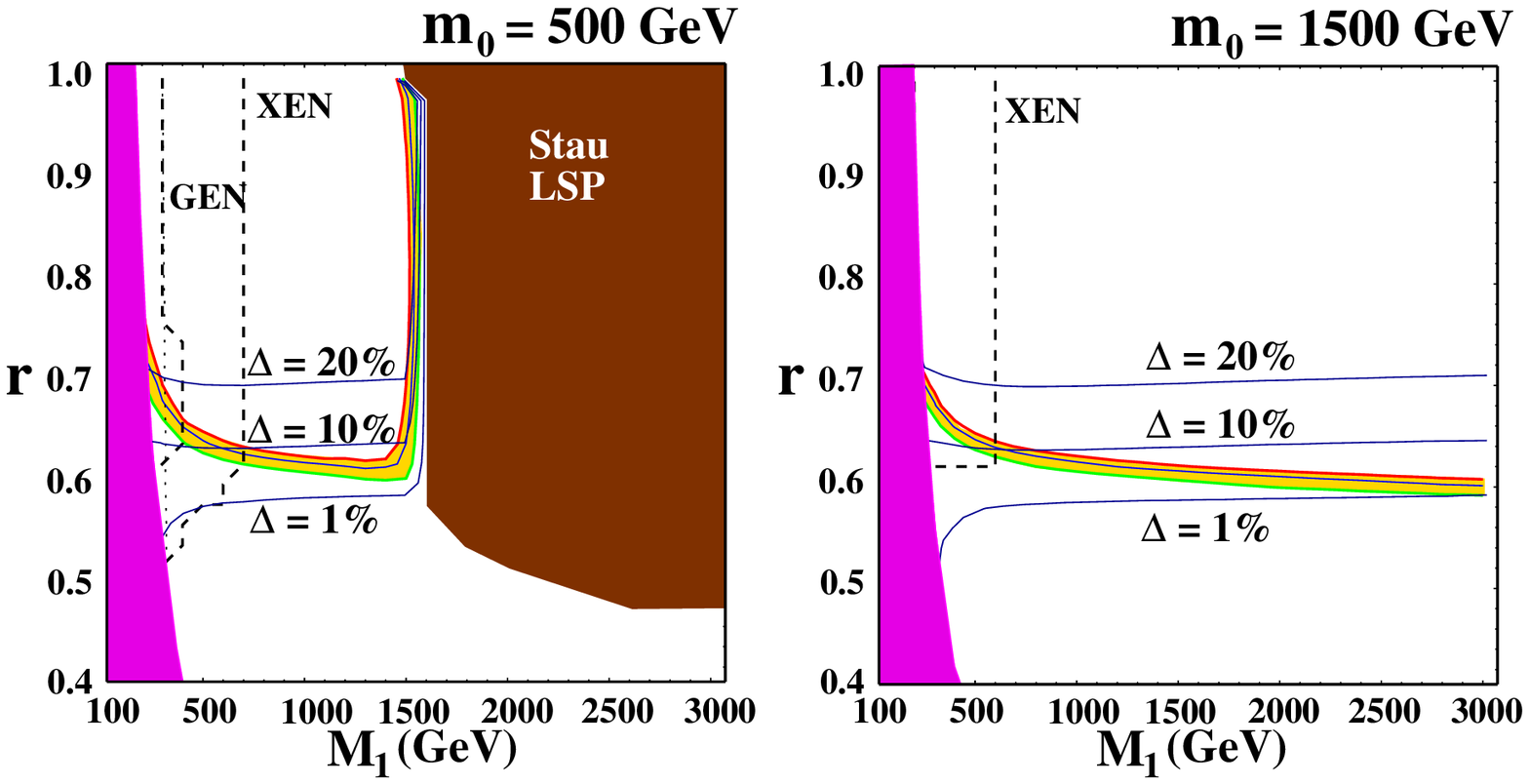}
  \includegraphics[scale=0.45]{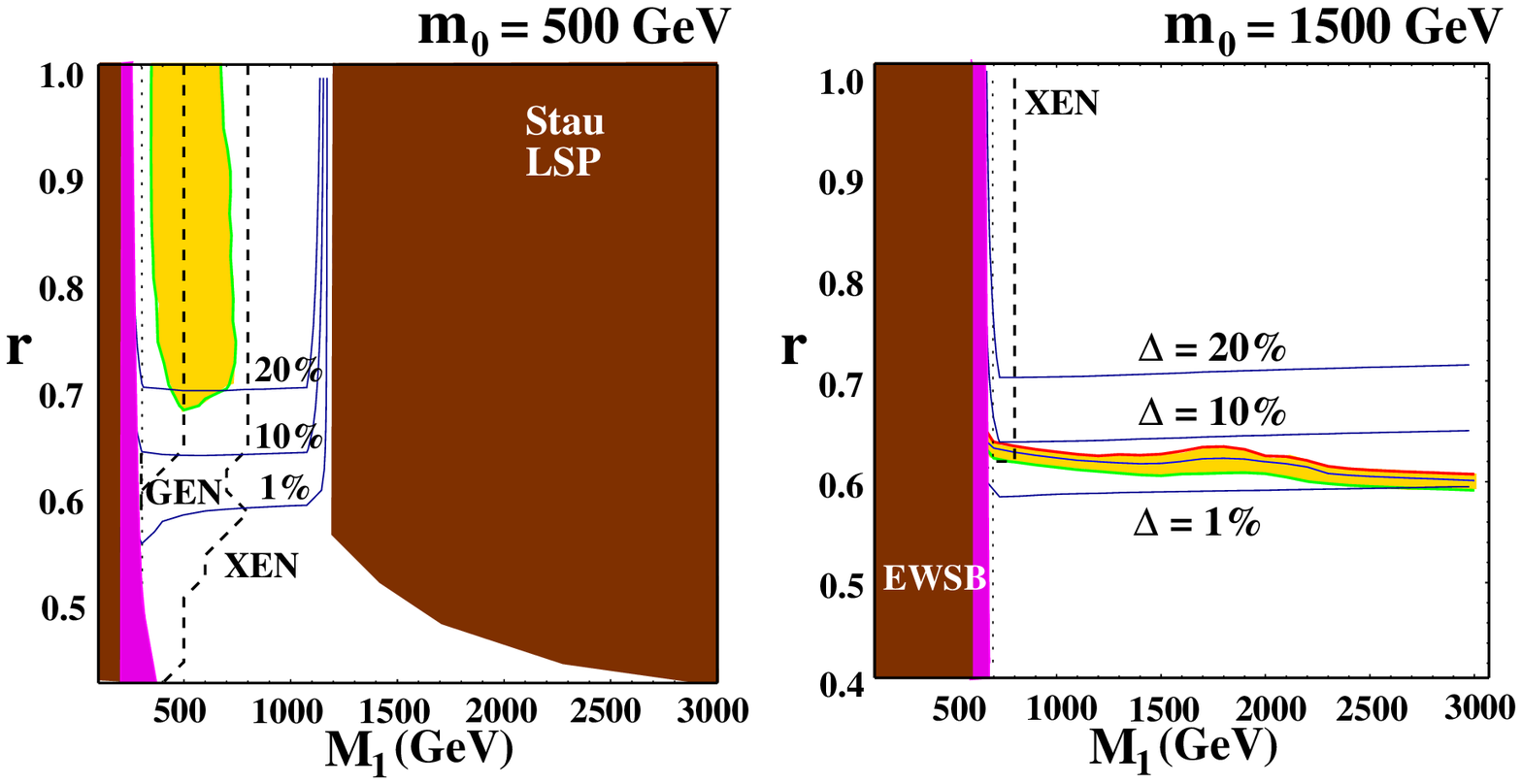}
  \caption{\label{fig:m_ra} {\bf Preferred Region in Gaugino Mass Space}. Top panels
            are for $\tan\beta=5$, middle panels are for $\tan\beta=35$ and
            bottom panels are for $\tan\beta=50$. The preferred neutralino relic
            density regions (light shading) are displayed for universal scalar mass
            $m_0 =500 \GeV$ (left panels) and for $m_0 = 1500 \GeV$ (right panels).
            The dark shaded region in the upper right of the left panels is excluded due
            to the charged LSP, while the region on the left in the $\tan\beta=50$ case
            is excluded by improper EWSB.
            The medium shaded regions on the left are excluded by LEP bounds on the
            chargino mass.
            The Higgs mass contour of $m_h = 113\GeV$ is given by the dotted vertical
            line.
            Contours of constant degeneracy parameter $\Delta$ are
            given by the thin near-horizontal lines.
            We have indicated by the dashed contours a rough estimate of the direct
            detection reach of GENIUS (``GEN'') and XENON (``XEN''), where applicable.}
\end{figure}

It would be tempting to conclude from Figures~\ref{fig:tan5}
through~\ref{fig:tan50} that relaxing the unification constraint
on the gaugino masses will increase the supersymmetric parameter
space that can properly account for the cold dark matter of the
universe. And while it is certainly true that any particular point
in the $\(M_1, \; m_0\)$ plane -- particularly those at larger
values of these soft parameters -- can now always accommodate the
cosmologically preferred relic density for some particular choices
of $r$ and $\tan\beta$, the size of the allowed areas in these
figures is somewhat deceptive. More information can be obtained by
exchanging the variable $m_0$ for $r$, since it is this parameter
that is more directly tied to the physics that determines the
eventual LSP relic density.

For this reason we plot the neutralino relic density in the
$\(M_1, \; r\)$ plane in Figure~\ref{fig:m_ra} for $\tan\beta=5$
(the top pair of plots), $\tan\beta=35$ (the middle pair of plots)
and for $\tan\beta=50$ (the bottom pair of plots). In this case we
have fixed the scalar mass to two different values to isolate the
effects of the gaugino sector parameters alone. Displayed in this
manner it is clear that the allowed region in gaugino mass space
is a narrow band that extends from the mSUGRA limit of $r=1$ in
the top left, where the region of direct annihilation between very
light LSPs is ruled out by the LEP chargino mass limits,
eventually extending to lower values of $r$, and then connecting
continuously with the stau coannihilation region for low values of
$m_0$. We believe that this space is the natural one for
discussing neutralino dark matter in nonuniversal gaugino mass
models: it conveys far more information on neutralino dark matter
than the typical $\(M_{1/2}, \; m_0\)$ plane in that the
correlation between degeneracy and relic density is readily
apparent.

To elaborate, we have plotted contours of constant degeneracy
degree $\Delta$ defined by
\begin{equation}
\Delta \equiv \frac{m_{\chi_{1}^{\pm}} -
m_{\chi_{1}^{0}}}{\bar{m}} ,
\label{delta} \end{equation}
where $\bar{m}$ is the average of the chargino and LSP masses. For
moderate to large values of the LSP mass, as determined by the
parameter $M_1$ at the GUT scale, the preferred region of
parameter space is in the neighborhood of $r\simeq 0.6$ where the
degree of degeneracy between the two mass eigenstates is roughly
$\Delta = 5\%$. In this region of the parameter space the degree
of degeneracy is just sufficient to allow coannihilation processes
between the chargino and LSP to sufficiently reduce the otherwise
large thermal relic density of the lightest
neutralino~\cite{BiNe01}. Clearly, as this degeneracy is increased
still further at lower values of the parameter $r$ the
coannihilation processes will be too efficient and the lightest
neutralino ceases to be an attractive dark matter candidate
without some method of non-thermal production. This is, for
example, the case in models of anomaly-mediation for which
$r_{\mathsc{amsb}} = M_2/M_1 = 5/33 \simeq 0.15$ at the GUT
scale~\cite{MoRa00}.

\begin{figure}
  \includegraphics[scale=0.45]{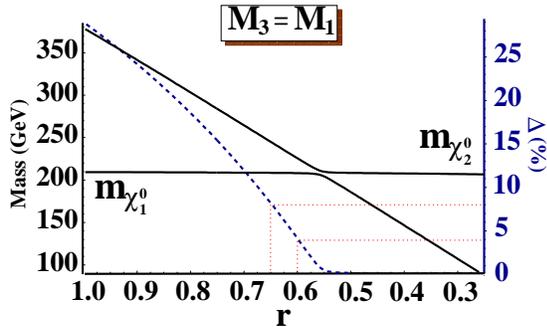}
  \caption{\label{fig:level1} {\bf Preferred Mass Degeneracy in the Gaugino System when
            $M_3 = M_1$ and $\tan\beta=35$}.
            The left vertical axis measures the mass (solid lines) of the lightest
            and next-to-lightest
            neutralinos $\chi_1^0$ and $\chi_2^0$, respectively, as a function of the
            parameter $r$.
            The right vertical axis measures the degree of degeneracy (dashed line)
            between the LSP
            and the chargino as given by the parameter $\Delta$, in percent, as a
            function of the same parameter.
            The cosmologically interesting region $0.65 \gappeq r \gappeq 0.60$
            is indicated by the red dotted lines.}
\end{figure}

The importance of this special value of $\Delta$, and hence of the
phenomenological parameter $r = M_2/M_1$, is no great mystery. As
was pointed out by Griest \& Seckel~\cite{GrSe91} the
coannihilating particle must be thermally accessible to the LSP
near the time of freeze-out for coannihilation processes to have a
significant impact on the overall relic density. Typical
freeze-out temperatures in the parameter space we consider are on
the order of $T_F \simeq m_{\chi_{1}^{0}} /20$, implying a
degeneracy on the order of a few percent is required to make
coannihilation processes relevant.

That the relic density should change so rapidly as $r$ is
diminished for fixed $M_1$ is also not surprising when one
considers the neutralino and chargino mass system. In
Figure~\ref{fig:level1} the mass of the lightest and
next-to-lightest neutralino are plotted as a function of the ratio
$r$ when $M_1$ is held fixed at $M_1 = 500 \GeV$ and $m_0$ is set
to $m_0 = 1000 \GeV$. The level crossing between a predominantly
B-ino LSP to a predominantly W-ino LSP occurs abruptly at $r
\simeq 0.55$. But just prior to this transition, near $r\simeq
0.65$ the degeneracy parameter $\Delta$ between the chargino and
the LSP drops rapidly through the cosmologically preferred region,
as is measured by the right vertical axis. Hence the
justification, {\em a posteriori}, for focusing on the case
$r=0.6$ in Figures~\ref{fig:tan5} through~\ref{fig:tan50}.

Of course, the nature of the gaugino mass system is also
determined by the value of the $\mu$ parameter at the electroweak
scale -- and in our procedure of fixing that parameter via the
EWSB conditions this implies some dependence on the gluino mass
$M_3$ and the various scalar masses through RG running. We
emphasize that the importance of the gluino mass on the relic
densities is of a secondary nature, through its indirect RGE
effects. Nevertheless, these effects can be significant in certain
regions of parameter space.

\subsection{\label{sec:M3M2}The case of gluino/w-ino equality}

In Section~\ref{sec:M3M1} we looked exclusively at the case where
$M_3 = M_1$ at the GUT scale in order to reproduce the typical RGE
behavior of universal models. In mSUGRA, with its unified gaugino
masses, most of the physics we associate with a given value of
$m_{1/2}$ is really the result of the RGE effects of the gluino.
When studying neutralino relic densities the value of $M_1$ at the
GUT scale is important in its own right, since the LSP mass is
typically $m_{\chi_1^0} \simeq 0.4 M_1$ in such scenarios. The
separate impacts of these two different mass scales are obscured
when $M_1 = M_3$, though this allows us to compare results to that
of mSUGRA more readily.

Rather than treat the gluino mass as a completely free variable we
prefer to keep the analysis tractable by invoking the relation
$M_3 = M_2$ in this subsection. Such a scenario will lead to
gluinos which are far lighter than in the previous section,
especially in the area around the critical region of $r\simeq
0.6$. What's more, we believe such a relation is likely to be
closer to the truth if gaugino mass nonuniversality arises in
models based on heterotic string theory.

For example, should gaugino mass nonuniversality arise at tree
level in weakly-coupled string models it would probably do so via
nonuniversal affine levels $k_a$ for the Standard Model gauge
groups ${\cal G}_a$, since gaugino masses at tree level are
determined by the auxiliary field of the dilaton which couples
universally to the gauge groups. Standard constructions for string
models enforce $k=1$ for non-abelian gauge groups such as $SU(3)$
and $SU(2)$, but generally have no such restriction on the affine
factor $k_Y$ associated with the $U(1)$ of
hypercharge.\footnote{See, for example, the discussion of this
point in Ref.~\onlinecite{Di97} and the references contained
therein.} Universality at tree level is recovered when the
phenomenological choice $k_Y = 5/3$, corresponding to the standard
GUT normalization of the hypercharge in the Standard Model, is
made. But this assumption is typically not borne out in explicit
string compactifications where exotic hypercharge normalizations
are quite common~\cite{DiFaRu96,ChHoLy96,ClFaSa01,Gi02}. While
such nonstandard hypercharge normalizations make gauge coupling
unification less straightforward, the presence of additional
states in the theory with exotic hypercharges can reconcile the
Standard Model gauge couplings with the universal coupling of the
underlying string theory~\cite{Gi02b}.

\begin{figure}
  \includegraphics[scale=0.45]{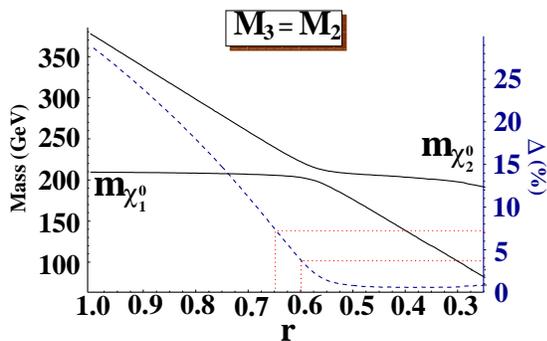}
  \caption{\label{fig:level2} {\bf Preferred Mass Degeneracy in the Gaugino System when
            $M_3 = M_2$ and $\tan\beta=35$}.
            The left vertical axis measures the mass (solid lines) of the lightest
            and next-to-lightest
            neutralinos $\chi_1^0$ and $\chi_2^0$, respectively, as a function of the
            parameter $r$.
            The right vertical axis measures the degree of degeneracy (dashed line)
            between the LSP
            and the chargino as given by the parameter $\Delta$, in percent, as a
            function of the same parameter.
            The cosmologically interesting region $0.65 \gappeq r \gappeq 0.60$
            is indicated by the red dotted lines.}
\end{figure}

\begin{figure}
  \includegraphics[scale=0.7]{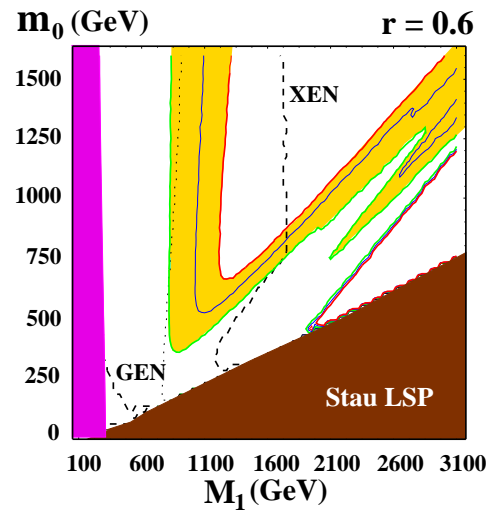}
  \caption{\label{fig:tan5b} {\bf Neutralino Relic Densities for
            $\tan\beta = 5$ and $M_3 = M_2$}. The preferred neutralino relic density
            region (light shading) is displayed for $r=0.6$.
            The shaded regions and contours are the same as those of
            Figure~\ref{fig:tan5}.}
\end{figure}

A perhaps more palatable alternative for nonuniversalities is that
the tree level gaugino masses are indeed universal, but somehow
suppressed through the mechanism of supersymmetry breaking and
transmission to the observable sector. Such a situation is, in
fact, quite common in realistic string theories and can arise when
the dilaton does not participate in supersymmetry breaking at all,
or when its lowest component is stabilized using stringy
nonperturbative contributions to its K\"ahler
potential~\cite{bench}. In both cases the resulting gaugino masses
are loop-suppressed, implying that non-universal corrections are
relevant. One class of terms arising at the one-loop level are
those associated with the superconformal
anomaly~\cite{GaNeWu99,BaMoPo00} which tend to increase the
gaugino mass $M_1$ far more than that of $M_2$ and $M_3$.

In such a regime where $M_2 \approx M_3$ the level crossing that
occurs in the neutralino system occurs far less abruptly. In
Figure~\ref{fig:level2} we again show the masses of the LSP and
NLSP neutralinos for $M_1 = 500 \GeV$, $m_0 = 1000 \GeV$ and
$\tan\beta=35$. The gluino mass at the GUT scale ranges from $500
\GeV$ at $r=1$ to $125 \GeV$ at $r=0.25$ on the right, leading to
much smaller values of the resulting $\mu$ term than in
Figure~\ref{fig:level1} and a smoother transition from B-ino like
to W-ino like LSP. Nonetheless, the critical region where $\Delta
\simeq 5\%$ continues to fall in the area near $r=0.6$.

The effect of the lighter gluino mass is seen most dramatically in
the plots analogous to those of
Figure~\ref{fig:tan5}-\ref{fig:tan50}. We return to the $\(M_1, \;
m_0\)$ plane for $\tan\beta=5$ in Figure~\ref{fig:tan5b}, where
now we have the relation $M_3 = M_2 = 0.6 M_1$. As in the right
panel of Figure~\ref{fig:tan5} the preferred cosmological region
is centered around $M_1 \simeq 1\TeV$, though this region is
narrower than in the previous case. More significant is the region
of rapid neutralino annihilation, and neutralino-chargino
coannihilation, through heavy Higgs states that now appears even
at low values of $\tan\beta$. These heavy Higgs eigenstates are
lighter than in a comparable unified model because of the
diminished effect of the gluino mass on the ultimate size of the
$\mu$ parameter, resulting in a smaller value than in mSUGRA. In
the case of Figure~\ref{fig:tan5b} two different resonant regions
can be resolved: the lower pole region being chargino-neutralino
resonant coannihilation through charged Higgs states, while the
upper pole region is neutralino-neutralino resonant coannihilation
through the heavy neutral Higgs states $H$ and $A$.

As mentioned earlier, it is well known that the size and location
in the $\(m_{1/2},\; m_0,\; \tan\beta\)$ space of this rapid
annihilation region is a strong function of the top and bottom
quark pole masses one assumes, as well as the treatment of the
relevant corrections to the heavy Higgs masses.\footnote{In fact,
even the choices of $\alpha_s(M_Z)$ and the exact definition of
the GUT scale $\Lambda_{\UV} = \Lambda_{\GUT}$ can have effects on
the pseudoscalar mass large enough to effect the position of this
depletion region. We have chosen the value $\alpha_s(M_Z) = 0.119$
and determine the high scale $\Lambda_{\UV}$ by the scale at which
$g_1$ and $g_2$ unify, usually close to the value of
$\Lambda_{\GUT} = 1.9 \times 10^{16} \GeV$.} The analysis of this
section suggests that on top of this sensitivity one should add
the sensitivity to the gluino mass, whose effects are felt through
the RGEs.

\begin{figure}
  \includegraphics[scale=0.45]{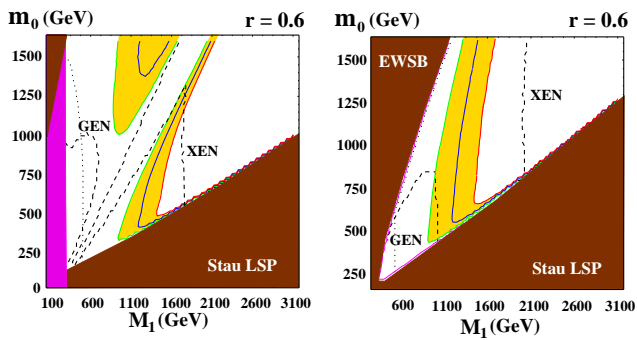}
  \caption{\label{fig:tan35b} {\bf Neutralino Relic Densities for
            $\tan\beta = 35$, $50$ and $M_3 = M_2$}. The preferred neutralino relic
            density regions (light shading) are displayed for $r=0.6$ and $\tan\beta = 35$
            (left panel) and $\tan\beta=50$ (right panel).
            The dark shaded regions in the lower right are excluded due to a charged LSP,
            while those in the upper left are excluded due to improper EWSB.
            The medium shaded regions on the left-most edges are excluded by LEP bounds
            on the chargino mass.
            The Higgs mass contour of $m_h = 113\GeV$ is given by the dotted vertical
            line.
            The dashed contours are a rough estimate of the direct
            detection reach of the GENIUS (``GEN'') and XENON (``XEN'') experiments.}
\end{figure}

This effect becomes even more pronounced at higher values of
$\tan\beta$, as in Figure~\ref{fig:tan35b}, where the heavier
Higgs boson states become lighter still. For the case of
$\tan\beta=35$ in Figure~\ref{fig:tan35b} (the left panel) the
resonant annihilation region has become quite pronounced as the
region where $2m_{\chi_1^0} \approx m_A ,\; m_H$ moves to lower
values of $M_1$. The size of the depletion region has also
increased substantially over the case where $M_3 = M_1$ and
$\tan\beta=50$. This is the result of resonant annihilation
through the pseudoscalar Higgs being possible at lower values of
$\tan\beta$, where the width $\Gamma_A$ is narrower and the
resonance annihilation effects are more pronounced. In the center
of this resonant annihilation region the neutralino relic density
is very low, implying a much lower count rate for GENIUS and XENON
and hence a lower direct detection reach in this area. By the time
$\tan\beta=50$ is reached (the right panel in
Figure~\ref{fig:tan35b}) this resonant annihilation region no
longer impacts the preferred relic density region and is confined
(for this value of $r=0.6$) to the region $M_1 \lappeq 1 \TeV$.
The ultimate conclusion one should draw is that the presence of
gluinos lighter than those in the universal case will generally
imply heavy Higgs resonant annihilation at lower values of
$\tan\beta$ than in mSUGRA, as well as shifting this region to
higher values of the universal scalar mass for fixed $M_1$ and
$\tan\beta$.

\begin{figure}
  \includegraphics[scale=0.45]{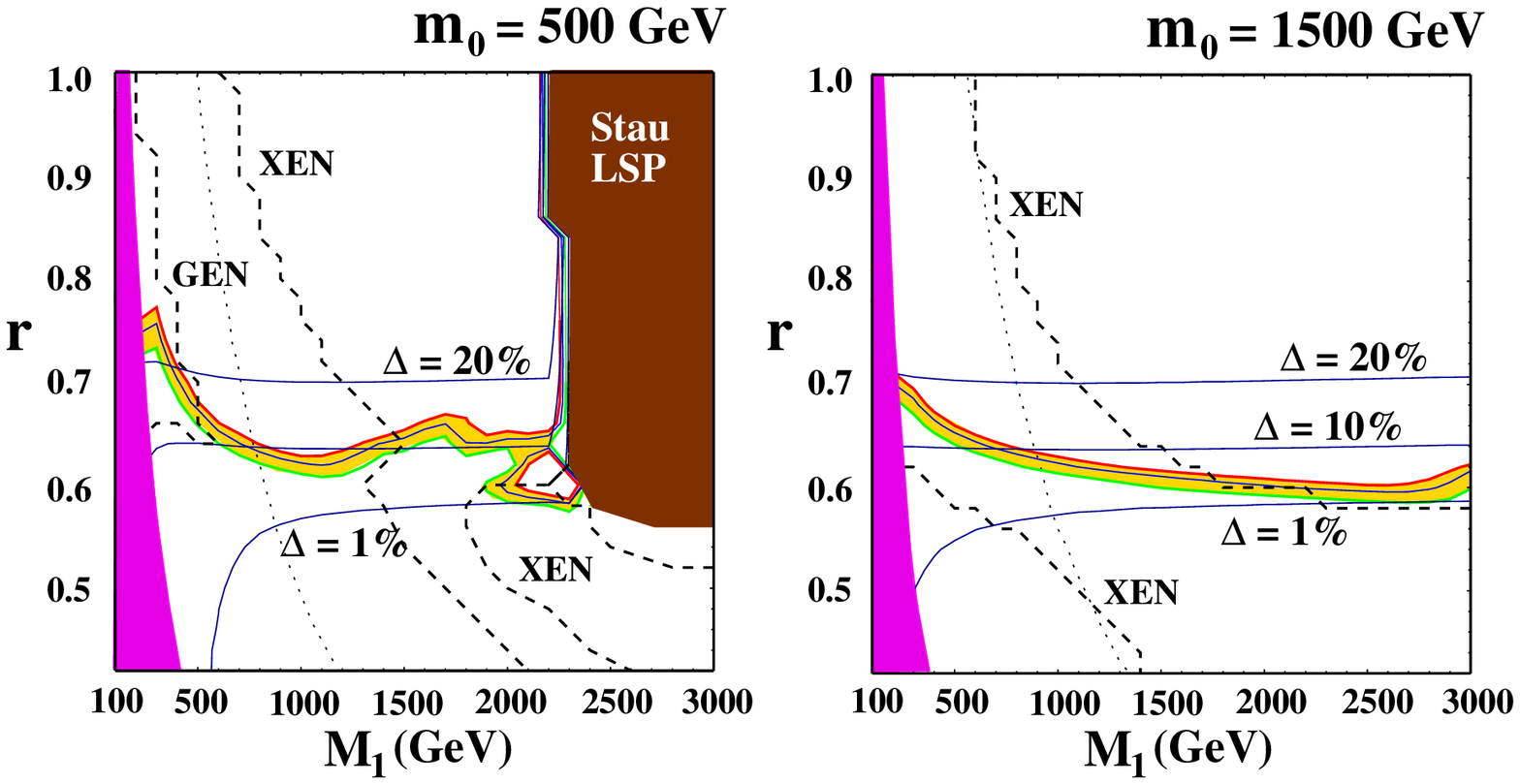}
  \includegraphics[scale=0.45]{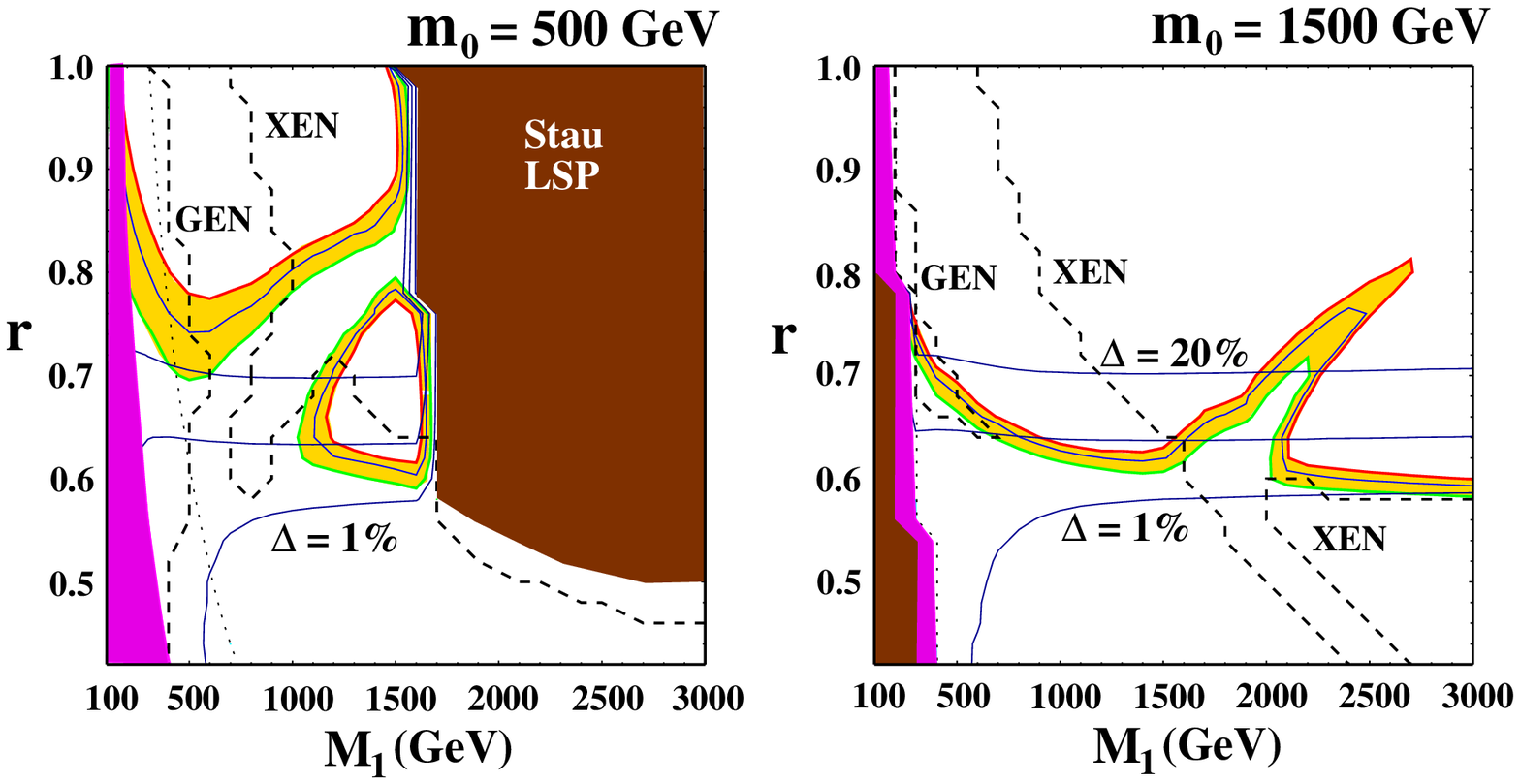}
  \includegraphics[scale=0.45]{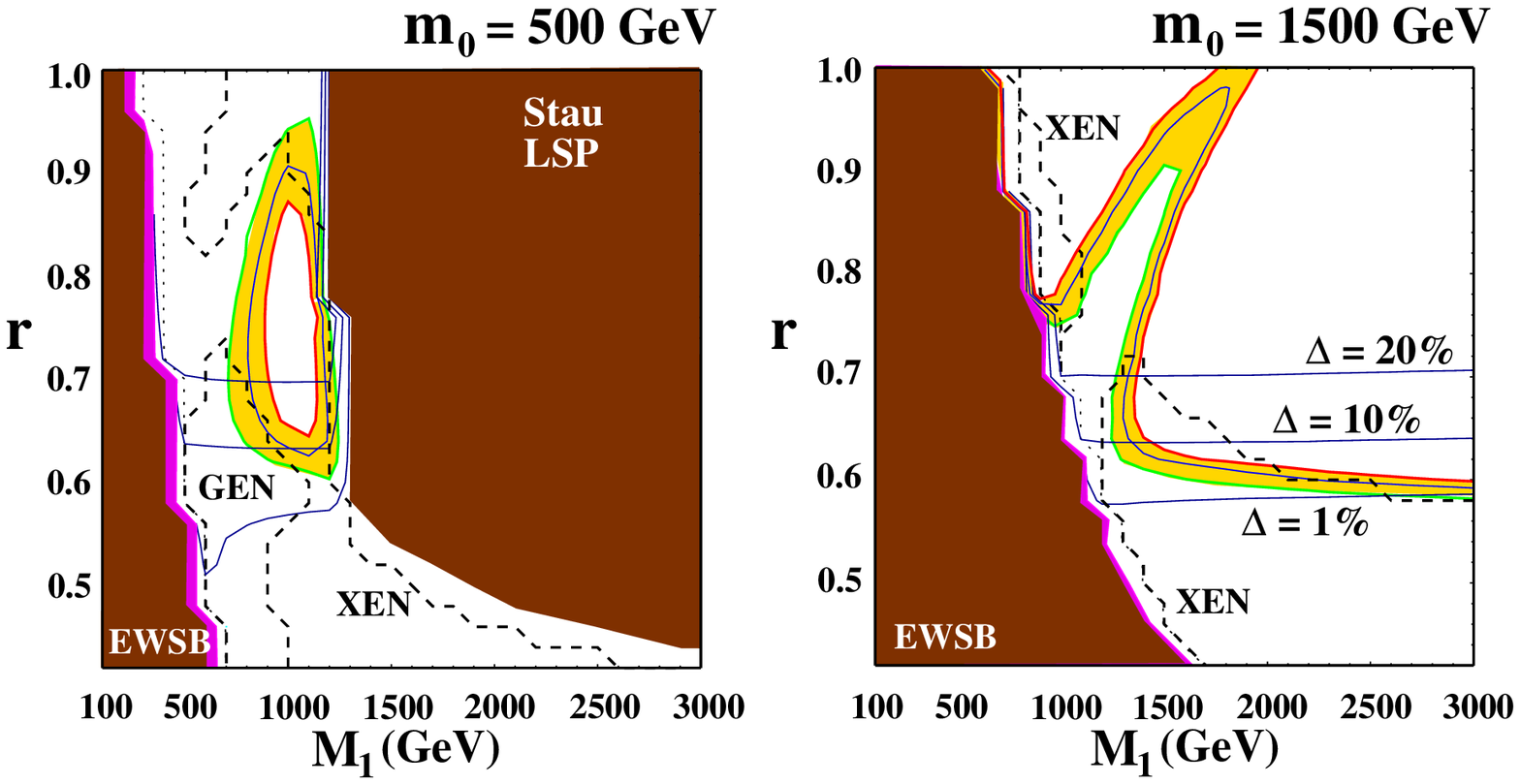}
  \caption{\label{fig:m_rb} {\bf Preferred Region in Gaugino Mass Space for
            $M_3 = M_2$}. Top panels
            are for $\tan\beta=5$, middle panels are for $\tan\beta=35$ and
            bottom panels are for $\tan\beta=50$. The preferred neutralino relic
            density regions (light shading) are displayed for universal scalar mass
            $m_0 =500 \GeV$ (left panels) and for $m_0 = 1500 \GeV$ (right panels).
            The contours in this figure are identical to those of Figure~\ref{fig:m_ra}.}
\end{figure}

It should be pointed out that by choosing to look in the $\(M_1,
\; m_0\)$ plane for the case where $r=0.6$ we have found that the
preferred region in that plane tends to fall around $M_1 =
\order(1\TeV)$ for both choices of $M_3$ and all three values of
$\tan\beta$. But this is an artifact of the choice $r=0.6$. The
LSP mass can be arbitrarily light, up to the constraint imposed by
the LEP bound on the chargino, by considering other values of $r$
for which chargino-neutralino coannihilations are less efficient.
This can be seen in Figure~\ref{fig:m_rb}, the analog to
Figure~\ref{fig:m_ra} shown previously, where lighter LSP
neutralinos can be obtained provided $0.7 \lappeq r \lappeq 0.9$,
depending on the value of $\tan\beta$. In this region the chargino
and neutralino are farther apart in mass so that $\Delta \gappeq
20\%$ and coannihilation processes are losing their relevance in
favor of the standard annihilation processes of typical mSUGRA
scenarios. This region of light charginos is also the region most
accessible to GENIUS and might be favored on general fine-tuning
arguments related to electroweak symmetry
breaking~\cite{KaLyNeWa02}.\footnote{These regions may also be
accessible to the proposed liquid-Xenon based ZEPLIN 4
detector~\cite{ZEPLIN} whose reach is somewhat less than the
GENIUS reach shown here.}

The distortions in Figure~\ref{fig:m_rb}, evident for low $m_0$ in
the $\tan\beta=5$ case and for both $m_0$ values in the higher
$\tan\beta$ cases, are a result of the resonant annihilation
region. In this region the correlation between chargino/neutralino
degeneracy, as measured by $\Delta$, and the neutralino relic
density is destroyed. Note also that the reduction in relic
density makes this region generally inaccessible experimentally,
even by the proposed XENON detector. When this distortion is
factored out the detectability reach in $M_1$ of relic neutralinos
in our local halo improves in the $M_3 = M_2$ regime over the case
where $M_3 = M_1$.

\section{\label{sec:detection}Prospects for Direct Detection of Relic Neutralinos}

We now wish to discuss the issue of direct detection of LSP
neutralinos in more detail, couching the discussion in terms of
the parameter space reach of the next-generation detectors such as
GENIUS and XENON. To compute the spin-independent
nucleon-neutralino elastic cross section the low-energy running
values of the soft supersymmetry-breaking terms at the scale
$M_Z$, as obtained by {\tt SuSpect}, were passed to the Fortran
code {\tt DarkSUSY}~\cite{DarkSUSY}.  The result of this analysis
is the spin-independent elastic cross section $\sigma_{\chi
p}^{\rm calc}$.

The ability of a given detector to observe a signal from relic
neutralino elastic scattering is a function of both the
interaction cross section $\sigma_{\chi p}^{\rm calc}$ and the
number density of LSPs in the local halo. Rather than plot event
rates per kilogram of target per year, we have chosen to follow
the custom in theoretical work on direct detection~\cite{BoFoSc01}
and plot an effective cross section as a function of the
neutralino mass $m_{\chi_1^0}$. This involves a rescaling of the
cross section, normalized to the case where the relic density in
the cosmos is the cosmologically preferred value. In our case we
have chosen the procedure outlined by Bottino et
al.~\cite{BoFoSc01}, which is to normalize the effective cross
sections whenever the relic density, as computed by {\tt
micrOMEGAs}, is below $(\Ohsq)_{\rm calc} = 0.1$ by the ratio $\xi
= (\Ohsq)_{\rm calc}/0.1$:
\begin{equation}
\sigma_{\chi p}^{\rm eff} = \xi \sigma_{\chi p}^{\rm calc} =
\(\frac{\Ohsq|_{\rm calc}}{0.1}\) \sigma_{\chi p}^{\rm calc} .
\label{rescale} \end{equation}
This procedure allows us to consider cases where the relic density
of neutralinos is far below the amount needed to provide all of
the cold dark matter of the cosmos. We feel such an approach is
important because (a) it represents a logical possibility given
that other sources of cold dark matter can be contemplated, (b) is
a more likely outcome in cases where $r <1$ than in the case of
mSUGRA, and (c) points in parameter space incapable of accounting
for all the needed cold dark matter are often points for which
direct detection rates are relatively high. Having performed this
rescaling we can now display model points together with the
expected reach in cross section for various dark matter detectors.
This is the manner in which the dashed detection reach contours
were computed for the figures in Section~\ref{sec:density}.

\begin{figure}
  \includegraphics[scale=0.45]{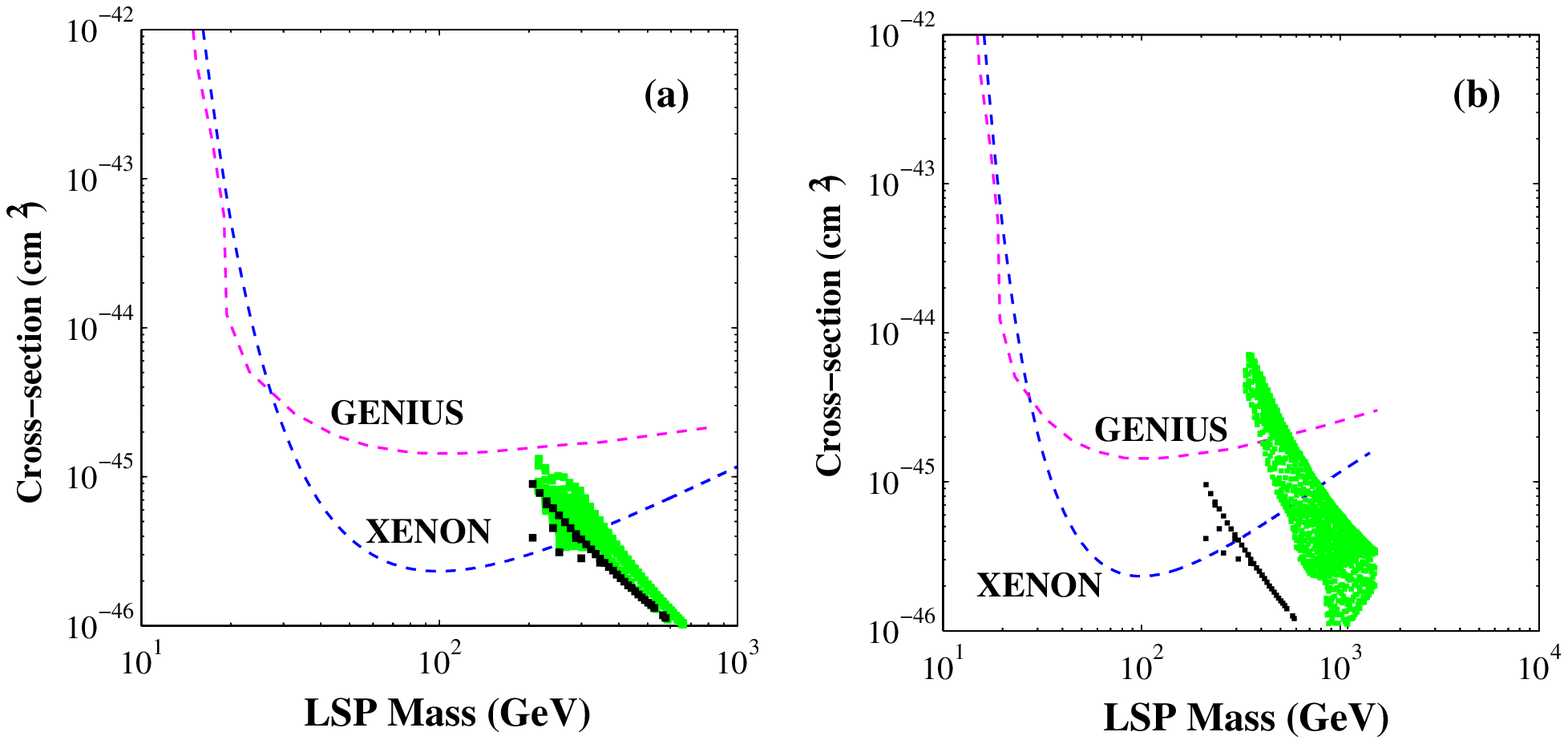}
  \includegraphics[scale=0.45]{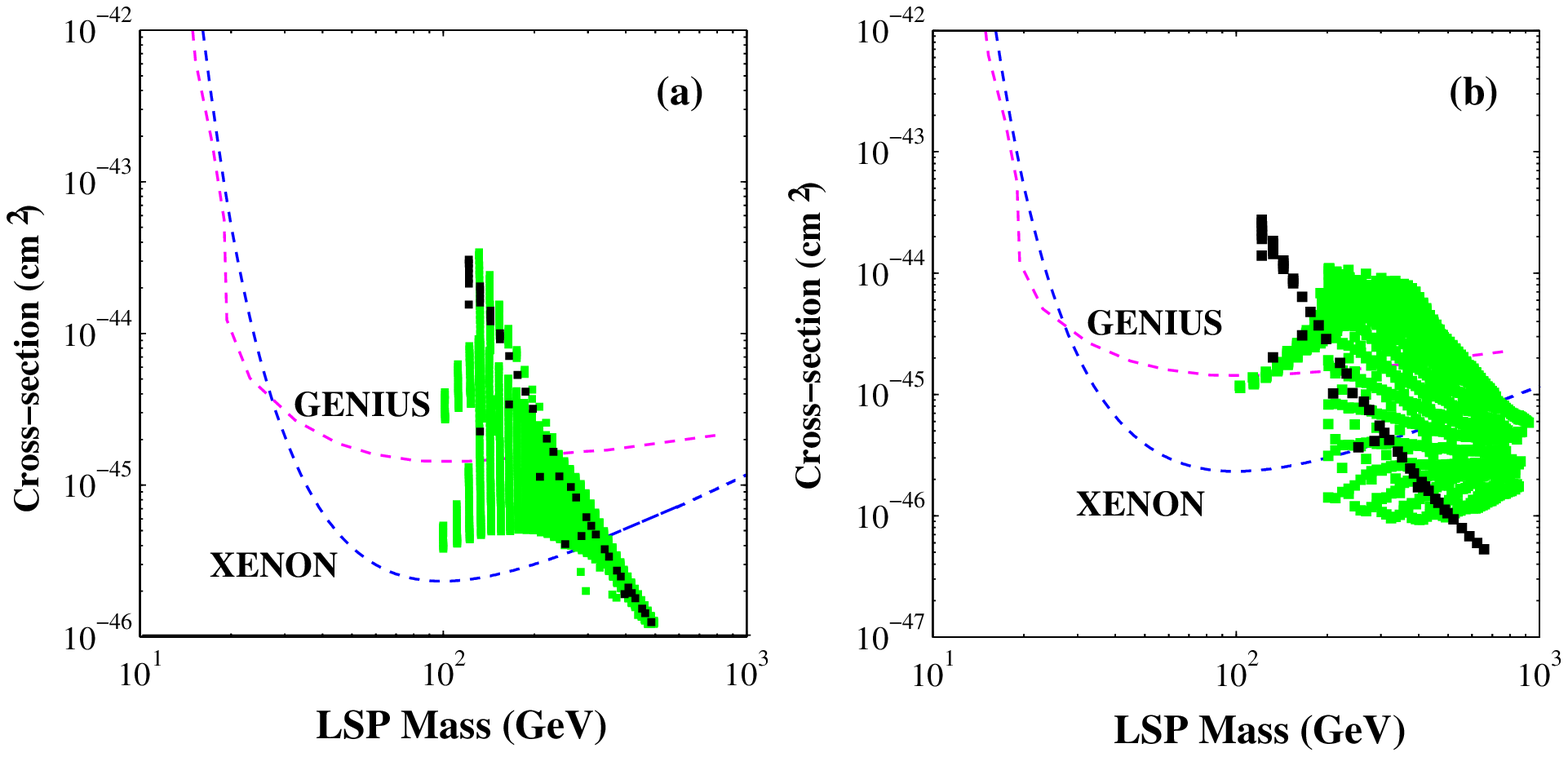}
  \includegraphics[scale=0.45]{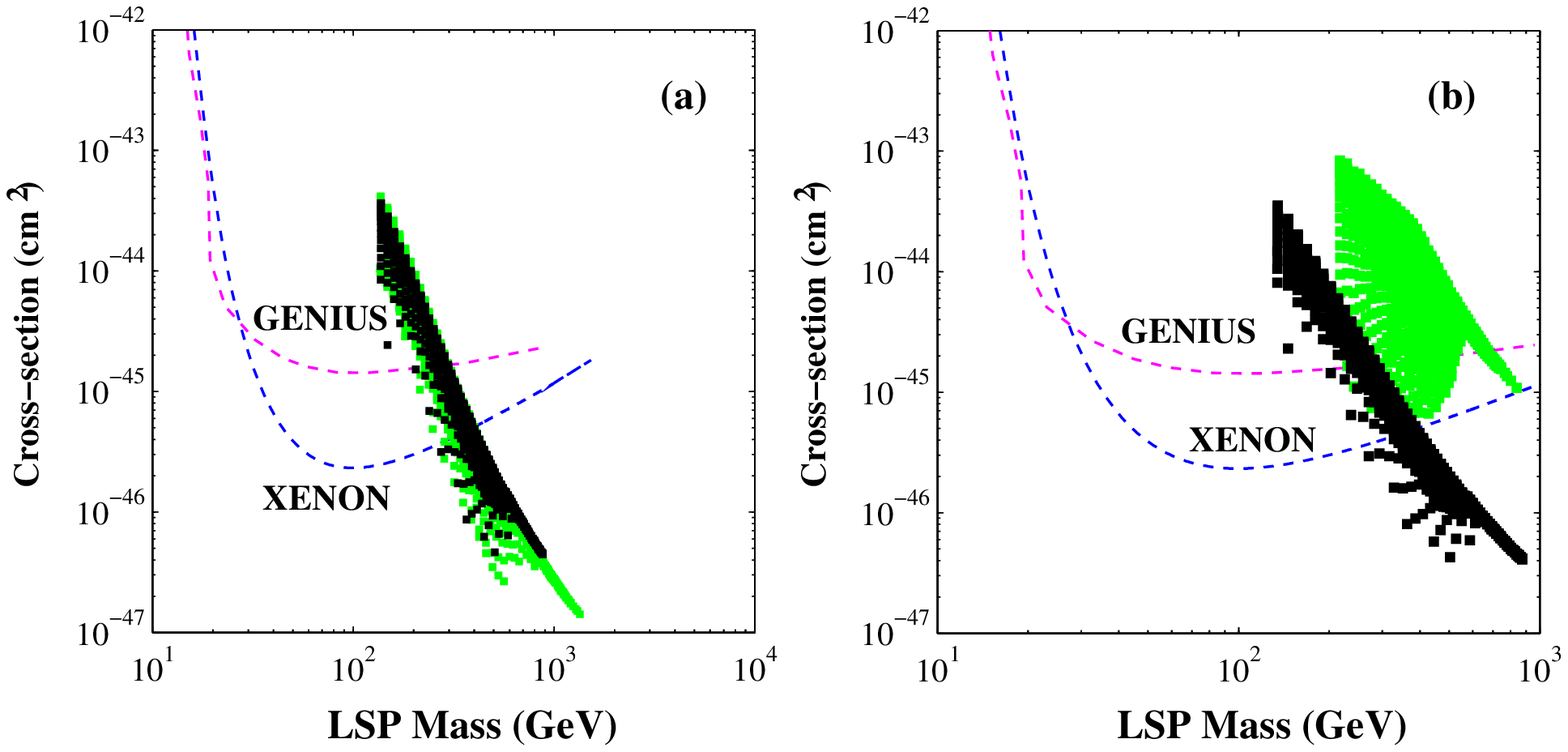}
  \caption{\label{fig:detect} {\bf Effective elastic cross sections
            and detector reach}. The rescaled, effective spin-independent cross
            section $\sigma_{p\chi}^{\rm eff}$ is plotted versus the physical LSP
            mass $m_{\chi_1^0}$ for $\tan\beta=5$ (top panels), $\tan\beta=35$
            (middle panels) and $\tan\beta=50$ (bottom panels).
            The left panels labeled (a) enforce the relation $M_3 = M_1$ of
            Section~\ref{sec:M3M1} while the right panels labeled (b) enforce the
            relation $M_3 = M_2$ of Section~\ref{sec:M3M2}.
            Dark (black) points are those of mSUGRA where $r=1$. The lighter (green)
            points are those in the region $0.60 \leq r \leq 0.65$.
            The estimation of the GENIUS and XENON detector reaches was obtained
            via the dark matter plotting program of Ref.~\onlinecite{dmplotter}.}
\end{figure}

A more common method for displaying the parameter space reach of a
particular detector is the manner of Figure~\ref{fig:detect}. In
this figure we have sampled the parameter space of
Figures~\ref{fig:tan5}-\ref{fig:tan50} and that of
Figures~\ref{fig:tan5b}-\ref{fig:tan35b} in intervals of $25\GeV$
for $m_0$ and $M_1$. We then eliminate those points that failed to
satisfy the EWSB and neutral LSP constraints, the LEP bounds on
the chargino and lightest CP-even Higgs masses, or had too large a
relic density $\Omega_{\chi} {\rm h}^2 > 0.3$. The calculated
elastic cross section is then rescaled, if necessary, to produce
the effective cross section $\sigma_{\chi p}^{\rm eff}$ which is
then plotted versus physical LSP mass. This procedure was
performed for $\tan\beta=5$ (the top panels), $\tan\beta=35$
(middle panels) and $\tan\beta = 50$ (lower panels). The left
panels labeled (a) enforce $M_3 = M_1$ while the right panels
labeled (b) enforce $M_3 = M_2$. The dark points are those of
mSUGRA in which $r=1.0$ while the lighter points include both the
case $r=0.60$ and $r=0.65$ to give some idea of the effect of the
gaugino mass nonuniversality on the typical effective
nucleon-neutralino elastic cross sections.

We should note that the reach of current and near-future direct
detection experiments does not extend to the effective cross
section range displayed in these plots. This is in accordance with
the usual result for models with unified scalar masses at some
large initial scale, such as mSUGRA, where the $\mu$ term tends to
be large compared to gaugino masses at the electroweak
scale~\cite{LaNaSp01,ElFeOl02,KiNiRoRu02}. Thus the LSP in all of
the cases we consider has a relatively small Higgsino content.
Also familiar from the unified mSUGRA case is the observation that
the typical size of the elastic scattering cross section increases
with higher $\tan\beta$. When the universality relation between
$M_2$ and $M_1$ is relaxed (while maintaining $M_3 = M_1$), as in
the left panels of Figure~\ref{fig:detect}, the increased
coannihilation cross section generally depresses the effective
cross section somewhat.

In the right panels of Figure~\ref{fig:detect} we tie the gluino
mass to $M_2$. In this case the lighter gluino results in lighter
squarks at the weak scale, as well as a lighter Higgs sector
though the smaller value of $\mu$ that results. This dramatically
increases the importance of those squark and Higgs exchange
diagrams that contribute to the spin-independent elastic cross
section for neutralinos on nucleons. The cross section in this
paradigm is augmented by one to two orders of magnitude: in
particular we note that at large $\tan\beta$ nearly the entire
range of parameter space shown will be covered by the proposed
XENON experiment. This marked increase in the elastic cross
section is consistent with a wide array of previous studies. The
reason is clear: any physics which reduces the size of the
low-energy squark and Higgs masses -- whether directly through
modifying the soft terms or indirectly through RG effects -- will
increase the importance of those squark and Higgs exchange
diagrams that contribute to the spin-independent elastic cross
section for neutralinos on nucleons. This includes the gaugino
mass nonuniversalities that lower $M_3$~\cite{BeNeOr02}, beginning
the RG evolution from a lower energy scale~\cite{GaKhMuTo01}, or
studying models with non-universal scalar
masses~\cite{ArDuHuSa01,KiNo01}.

\section{\label{sec:tuning}Fine-Tuning and Sensitivity Analysis}

The question of fine-tuning in a given parameter, or set of
parameters, comes down to asking how natural it is to end up with
the parameter at a certain value. Such a determination should
ultimately be done in the context of some underlying theory, such
as string theory. Without the luxury of such a fundamental theory
we are forced to rely on constructed notions of fine-tuning. The
area of fine-tuning has attracted many different definitions and
interpretations.\footnote{See, for example, the discussion of this
point in Ref.~\onlinecite{FeMa01}.} We are not especially fond of
any particular measure of tuning. Nonetheless, we think that a
firm, standard definition should be used when comparing models and
regions of parameter space. There are two common measures of
fine-tuning employed when discussing relic densities.

First is a frequent, though soft, definition of fine-tuning we
might call the ``optical'' or ``likelihood'' method.  Here one
simply looks at a plot of parameter space and ``eye-balls''
whether or not the allowed parameter space is fine-tuned by
estimating the relative sizes of the preferred volume of parameter
space to the entire volume otherwise allowed in the theory. For
instance, looking at the $\tan\beta=5$ plot in
Figure~\ref{fig:tan5} for mSUGRA (the $r=1$ limit for the rSUGRA
models), one notices that the relic density depends on a very
specific ratio of $m_{0}$ to $M_{1/2}=M_1$. Since only a very thin
sliver of parameter space meets the relic density constraint, one
would conclude, using the likelihood method, that dark matter is
fine-tuned in mSUGRA for $\tan\beta=5$. As we will see below, this
is not necessarily the case when more robust measures of
fine-tuning are employed.

The optical method can be used on the rSUGRA parameter space to
come to any desired conclusion about fine-tuning.  For instance,
looking at the $M_{1}$ vs. $m_{0}$ plot for $r=0.6$ in
Figure~\ref{fig:tan5}, the preferred dark matter region does not
look especially tuned: a wide region exists that meets the relic
density constraint. However, the opposite conclusion is reached if
one looks instead at the $M_1$ vs. $r$ plot for $\tan\beta=5$ in
Figure~\ref{fig:m_ra}. As noted earlier, on this plot the
parameter space looks quite narrow and is obviously rather
dependent on $r$.

To eliminate this ambiguity, in this paper we choose the second
and slightly less subjective definition commonly employed, which
is based on a ``sensitivity'' measure~\cite{ElEnNaZw86,BaGi88} as
applied to neutralino relic densities by Ellis and
Olive~\cite{ElOl01}
\begin{equation}
\Delta_{\Omega, i} \equiv \frac{a_{i}}{\Ohsq}\frac{\partial\(
\Ohsq\)}{\partial a_{i}} .
\label{tuning} \end{equation}
The $a_{i}$ in~(\ref{tuning}) represent parameters that are to be
varied in the theory, typically soft breaking parameters though
sometimes also the top and bottom Yukawa couplings or masses. Even
though this definition is not without limitations~\cite{AnCa95},
we find it at least provides a common language which makes
comparisons easier. The overall level of fine-tuning then is given
by
\begin{equation}
\Delta_{\Omega} =\sqrt{\sum_{i} (\Delta_{\Omega, i})^2}.
\label{sumDelta} \end{equation}

We will compute two different $\Delta_{\Omega}$'s for certain
sample points in the generalized rSUGRA parameter space.  In the
first we compute $\Delta_{\Omega}$ for the case where the $a_{i}$
belong to the set $\{r,\; m_{0},\; M_{1},\; \tan\beta,\; m_{t},
\;m_{b}\}$. Since we will restrict ourselves to one of the gluino
mass relations of Sections~\ref{sec:M3M1} and~\ref{sec:M3M2}, this
parameter set represents the maximal sensitivity measure for a
given point. But we will also consider the case where a more
fundamental theory predicts one specific value of $r$, so we will
also compute the quantity $\Delta_{\Omega}^{r \!\!\! \not}$, where
the $a_{i}$ now belong to the smaller set $\{m_{0},\;
M_{1},\;\tan\beta,\;m_{t}, \;m_{b}\}$. For example, the variation
of $r$ in mSUGRA has no meaning, since $r=1$ is one of the
defining characteristics of this model.\footnote{Of course the
most well-studied unified model to lay claim to a string theory
origin -- namely that of the dilaton domination model at
tree-level -- predicts even more constraints among the various
soft parameters of the mSUGRA paradigm.} Thus we will only compute
the quantity $\Delta_{\Omega}^{r \!\!\! \not}$ for the mSUGRA
scenario as a baseline for comparison with the expanded parameter
space of rSUGRA.

We begin our sensitivity analysis by considering the mSUGRA case
as a baseline.  In Table~\ref{tbl:mSUGpref} we provide the
sensitivity parameter $\Delta_{\Omega}^{r \!\!\! \not}$ for a
sample of points which yield the cosmologically preferred relic
density (the shaded areas in the left panels of
Figures~\ref{fig:tan5},~\ref{fig:tan35} and~\ref{fig:tan50}).
These values are in very good agreement with those calculated in
Ref.~\onlinecite{ElOl01} for similar points in the parameter
space. The sensitivity is greatest for large $\tan\beta$, which is
expected due to the presence of the resonant annihilation area
which is itself sensitive to $\tan\beta$ and $m_{b}^{\rm pole}$.

Let us compare these typical values with those of various points
in the rSUGRA parameter space. Table~\ref{tbl:rSUGpref} looks at
the sensitivity parameters for a set of points under three
different gaugino mass relations: $M_{3}=M_{1}$, $M_{3}=M_{2}$,
and $M_{2}=M_{1}$. The first two represent the case of
Sections~\ref{sec:M3M1} and~\ref{sec:M3M2}, respectively, while
the third enforces $M_{3} = rM_1 = rM_2$. This third relation was
studied by the authors of Ref.~\onlinecite{BeNeOr02} and we
include it here to make contact with their observations.
For these rSUGRA scenarios we have chosen points which have the
correct relic density in the $M_{3}=M_{2}$ case
(Figure~\ref{fig:m_rb}), but many of the points also have the
correct relic density in the $M_{3}=M_{1}$ case so the comparison
is still instructive.

\begin{table}[t]
\caption{\label{tbl:mSUGpref} Relic Density Sensitivity for
Preferred Points in mSUGRA Parameter Space.}
{\begin{center}
\begin{tabular}{|ccc||c|}
\cline{1-4}
$M_{1}$ & $m_{0}$ & $\tan\beta$ & $\Delta_{\Omega}^{r \!\!\!
\not}$ \\ \cline{1-4}
 1100 & 235 & 5 & 13.0\\
 1500 & 343 & 5 & 15.7\\
 1100 & 390 & 35 & 31.8\\
 1100 & 550 & 50 & 78.3\\
 1100 & 950 & 50 & 124.1\\
 1600 & 740 & 50 & 44.2\\
 1600 & 1500 & 50 & 131.9\\
\cline{1-4}
\end{tabular}
\end{center}}
\end{table}

\begin{table}[t]
\caption{\label{tbl:rSUGpref}Relic Density Sensitivity for
Preferred Points in rSUGRA Parameter Space.}
{\begin{center}
\begin{tabular}{||cccc||cc||cc||cc||}
%
\multicolumn{4}{c}{Point} & \multicolumn{2}{c}{$M_{3}=M_{1}$} &
\multicolumn{2}{c}{$M_{3}=M_{2}$} &
\multicolumn{2}{c}{$M_{2}=M_{1}$} \\ \cline{1-10}
$M_{1}$ & $m_{0}$ & $\tan\beta$ & $r$ &$\Delta_{\Omega}$ &
$\Delta_{\Omega}^{r \!\!\! \not}$ & $\Delta_{\Omega}$ &
$\Delta_{\Omega}^{r \!\!\! \not}$ & $\Delta_{\Omega}$ &
$\Delta_{\Omega}^{r \!\!\! \not}$ \\ \cline{1-10}
 900 & 350 & 5 & 0.60 & 50.6 & 3.3 & 39.2 & 15.4 & 6.6 & 4.5\\
 1100 & 1500 & 5 & 0.60 & 54.0 & 3.4 & 50.1 & 7.7 & 5.4 & 5.2\\
 3000 & 1500 & 5 & 0.60 & 48.9 & 2.5 & 40.7 & 16.3 & 250.2 & 218.6 \\
 3000 & 1375 & 5 & 0.60 & 48.8 & 2.5 & 49.4 & 39.4 & 53.6 & 47.5 \\
 3000 & 1250 & 5 & 0.60 & 48.7 & 2.5 & 86.1 & 73.5 & 56.2 & 49.7 \\
 1100 & 1500 & 35 & 0.60 & 53.5 & 3.3 & 47.6 &17.2 &70.7 &69.8\\
 1100 & 1000 & 35 & 0.60 & 50.0 & 2.5 & 73.6 &68.9 &70.9 &68.8\\
 1600 & 1500 & 35 & 0.60 & 51.5 & 2.6 & 80.7 &74.4 &115.5 &112.7\\
 2100 & 1500 & 35 & 0.60 & 47.8 & 2.7 & 24.0 &6.0 &58.2 &57.8\\
 400 & 500 & 35 & 0.67 & 70.8 & 62.9 & 22.5 & 22.3 & 136.1 & 136.0\\
 300 & 1500 & 35 & 0.75 & 9.9 & 9.9 & 20.3 & 18.9 & 18.0 & 17.4\\
 1100 & 750 & 50 & 0.60 & 27.8 & 23.1 & 26.8 & 6.0 & 8.4 & 7.8\\
 1600 & 1500 & 50 & 0.60 & 39.8 & 32.7 & 28.6 & 7.2 & 19.7 & 18.8\\
\cline{1-10}
\end{tabular}
\end{center}}
\end{table}

One can see that when $r$ represents the ratio of $M_2$ to $M_1$,
as it does in the first two columns of Table~\ref{tbl:rSUGpref},
its inclusion in the set $\{a_{i}\}$ has an important effect on
the overall degree of sensitivity. This is especially striking in
the first column in cases where the LSP is quite massive (large
$M_1$) where the correct relic density is obtained almost
exclusively through coannihilation effects. Here the sensitivities
are everywhere comparable to those of mSUGRA in
Table~\ref{tbl:mSUGpref}. The large impact of including $r$ in the
set $\{a_{i}\}$ is merely verifying that this one parameter
effectively captures the key physics involved in the relic density
calculation.

In the second and third column, where the gluino mass is allowed
to vary and is generally smaller than in the first column, the
impact of the variable $r$ is somewhat muted. As mentioned
previously, the impact of $M_3$ is on the lower squark masses and
the lower value of $\mu$. In the extreme case of the third column
of Table~\ref{tbl:rSUGpref}, where $M_2= M_1$ and only the gluino
mass is allowed to vary (downwards) we see that the impact of the
relative size between $M_3$ and the other gaugino masses is minor.
What is far more important is the absolute magnitude of $M_3$,
particularly in points associated with the region near a Higgs
resonance. In these examples the fine tuning is comparable to that
of mSUGRA near a Higgs resonance and is $\order(100)$ in size.
This indicates that when resonant annihilation processes are
important it is the gluino mass at the high scale that captures
the largest fraction of the physics involved.

Thus, in the worst case scenario for $M_{2}$-floating rSUGRA, the
relic density sensitivity is a factor of $4$ larger than a
comparable point in mSUGRA parameter space.  In the best case
scenario, rSUGRA relic density sensitivities are {\it better} than
comparable fine-tunings in mSUGRA parameter space by a factor of
$4$. This is true even ignoring the possibility that $r$ might not
be included in the set of variation parameters.  If $r$ is
excluded, the relic density fine-tunings of rSUGRA compare even
more favorably with those of mSUGRA.

So what are we to conclude from Tables~\ref{tbl:mSUGpref}
and~\ref{tbl:rSUGpref}? On the one hand we agree with the authors
of Ref.~\onlinecite{BeNeOr02}: achieving the preferred
cosmological relic density requires that the parameter $r$ be
tuned to a specific narrow range for any given universal scalar
mass $m_0$, B-ino mass $M_1$ and $\tan\beta$. But the same could
be said for any model that predicts a stable relic neutralino. In
mSUGRA one must tune $\tan\beta$ and/or the mass ratio of the stau
and the lightest neutralino to a specific narrow range to get the
right answer. In cases where the gluino is separated from the
(still unified) B-ino and W-ino masses one must tune {\em its}
value to a narrow range in order to achieve just the right value
of $\mu$ at the electroweak scale. Tuning of this type is common
to phenomenology where we lack a fundamental theory that can {\em
explain} such relations.

On the other hand we also agree with the authors of
Ref.~\onlinecite{ElOl01}: these tunings that are required of the
theory are generally speaking not {\em fine}-tunings. That is to
say, the sensitivity to input parameters represented by the values
for $\Delta_{\Omega}$ in Tables~\ref{tbl:mSUGpref}
and~\ref{tbl:rSUGpref} are on the whole quite mild. In fact the
relatively small tuning parameters $\Delta_{\Omega}$ are all the
more surprising when we recall that calculating the relic density
in the presence of coannihilations involves the exponential of the
mass difference $\Delta$ of~(\ref{delta}). If a stable neutralino
is a significant component of the cold dark matter of the cosmos
(and we reiterate that it certainly need not be), then some
specific area of the MSSM parameter space will inevitably be
singled out. But it is worth bearing in mind that the tuning of
parameters implied in Tables~\ref{tbl:mSUGpref}
and~\ref{tbl:rSUGpref} is generally small in comparison with the
tuning required in the electroweak symmetry-breaking sector to
obtain the correct $Z$-boson mass -- a physically measured
quantity.

\section*{Conclusion}

Rapidly advancing calculational tools have put the computation of
many important observables relating to relic neutralinos at hand
to an ever-widening community. We believe making sound judgements
as to the region of supersymmetric model space favored by
cosmological observations on the density of cold dark matter
requires utilizing these tools in directions beyond that of
unified models. In this work we have focused on the case of
nonuniversal gaugino masses and, in particular, a paradigm in
which the W-ino mass $M_2$ (and perhaps the gluino mass) are
significantly smaller at the high-energy input scale than the
B-ino mass $M_1$. Such a regime may be a likely outcome of
effective supergravity theories based on the weakly-coupled
heterotic string, though we have presented the implications of
these nonuniversalities in a model-independent manner.

Many complications and uncertainties exist that should give us
pause in trying to make exact quantitative statements: broad
theoretical questions such as whether there may be non-thermal LSP
production mechanisms or whether phase transitions at late times
may have generated some degree of late-term inflation. There are
also technical issues that can have a significant impact on one's
results, particularly at large $\tan\beta$ where the choice of
physical pole masses for the quarks, the accuracy of the RG
evolution employed, and the completeness of the corrections to
Higgs masses and widths are all crucial.

Nonetheless, it is clear that relaxing the unification condition
on gaugino masses opens up new directions in the parameter space
which are every bit as theory-motivated and phenomenologically
appealing as those of mSUGRA. These new directions may improve the
prospects for detecting relic neutralinos from our halo in future
experiments, particularly the proposed liquid Xenon experiments.
The necessary tunings involved may, in fact, be a tool in helping
us to ascertain the correct underlying theory once a discovery is
made (for example, selecting a narrow list of possible hidden
sector condensing gauge groups in string-derived supergravity
models~\cite{BiNe01}). In this sense the work presented here on
nonuniversal gaugino masses forms an attractive complement to the
recent work on nonuniversal Higgs
masses~\cite{ElFaOlSa02,ElOlSa02}. Both are well-defined
departures from the baseline mSUGRA paradigm which can claim solid
motivation from heterotic string inspired models. Computation of
the relic density and elastic scattering cross section for relic
neutralino LSPs is a wonderful laboratory for studying
supersymmetric phenomenology in that it combines elements from the
squark, slepton, gaugino and Higgs sectors in inter-related ways.
We believe disentangling these diverse physics contributions is
best performed with a toolkit that contains a variety of more
general MSSM models with which to work.

\begin{acknowledgments}
The authors would like to thank M.~K.~Gaillard for helpful
conversations and suggestions. BDN would like to thank the
Lawrence Berkeley National Laboratory for hospitality during the
early portions of this work.
\end{acknowledgments}


\end{document}